# A MATHEMATICAL MODEL OF MOTIVATED EMOTIONAL MIND - COGNITIVE EMBODIED SYSTEM

Wiesław L. Galus, and Janusz A. Starzyk

***Abstract*****—**This article presents a mathematical model of the Motivated Emotional Mind cognitive architecture developed for embodied intelligent systems. Such a system learns to maintain its homeostasis through a generalized form of reinforcement learning based on its internal motivations, termed motivated learning (ML). The principal contribution of this article is a rigorous formalization of the re-entrant loop integrating feedforward processing, lateral interactions, and feedback pathways, together with the representational selection mechanisms that govern adaptive system responses. The model specifies how ongoing exteroceptive and interoceptive signals, bodily-motivational context, and memory traces are bound into associative memory structures termed semblions, which compete for access to further processing and top-down reconstruction. The formalization encompasses secondary perception, representational competition, curiosity, procedural gaps, and action selection directed toward limiting allostatic violations. Within this framework, motivated learning is tailored to embodied systems whose dynamics are shaped by needs, affect, and the current regulatory state. Unlike standard reinforcement-learning models, the proposed approach incorporates need thresholds, goal generation and shifting goals, bodily state, resource constraints, and action uncertainty, thereby providing a more adequate account of response selection under regulatory pressure. Global affect functions as a central control signal, modulating the learning rate, representational valence, and the balance between exploration and exploitation. The model presented here is a step toward a more rigorous formalization of cognitive phenomena and may provide a basis for further theoretical analysis, computer simulation, and implementation in artificial-intelligence systems inspired by biological processes.

## 1. INTRODUCTION

A mathematical model of phenomenal consciousness must do more than attach variables to familiar psychological terms. It must describe subjectivity through measurable processes: receptor-driven stimulation, bodily regulation, affective valuation, memory-based recognition, and re-entrant reconstruction of sensory and interoceptive maps. The central thesis of this paper is that phenomenal consciousness is constituted by receptor-grounded, affectively modulated, recurrent reconstruction of sensory and interoceptive representations in an embodied system. This hypothesis is formalized within the proposed Motivated Emotional Mind (MEM) framework. Its aim is to explain how first-person experience may arise in a biological or biologically inspired system without invoking a non-physical substance, primitive informational property, or panpsychist proto-experience.The philosophical pressure behind such a model was sharpened by twentieth-century debates. Ryle (1949) rejected the Cartesian "ghost in the machine," while Place (1956), Feigl (1958), and Smart (1959) identified mental states with brain processes. Yet identity theory did not specify which neural processes make sensing a sensation. Nagel's (1974) question, Levine's (1983) explanatory gap, and Chalmers's (1995, 1996) hard problem exposed the same missing link: a physical theory must explain why certain activations constitute felt qualities. MEM accepts this challenge but treats the gap as evidence for a missing mathematical and biophysical account.

Psychophysics was the first systematic attempt to mathematize the relation between stimulation and conscious appearance. Weber and Fechner sought lawful relations between stimulus magnitude and sensation, and Fechner's Elements of Psychophysics treated the body–mind relation as an exact science (Fechner, 1860/1966). Stevens (1957) later introduced power-law scaling, and signal detection theory separated sensitivity from decision criterion (Green & Swets, 1966). Although these approaches did not solve the problem of qualia, they established a methodological standard: subjective experience becomes scientifically tractable only when tied to formal variables and experimentally controllable processes.

Cybernetics, information theory, and control theory provided a second line of formalization. Shannon (1948) supplied the mathematics of signal transmission, while Wiener (1948) and Ashby (1952) treated adaptive behavior as regulation under feedback. This tradition encourages a temptation MEM resist: identifying consciousness with information processing itself. Information can be measured without being felt, and a controller can minimize error without pain. Thus, information, feedback, and optimization are necessary but insufficient for a model of consciousness, unless embedded in a body with receptors, needs, affective pressure, and re-entrant sensory constitution.

Contemporary theories contributed further formal ingredients. Global Workspace Theory modeled conscious access as competition and broadcast (Baars, 1988; Dehaene & Changeux, 2011). Higher-order theories stressed representations of representations (Rosenthal, 2005; Lau & Rosenthal, 2011). Integrated Information Theory introduced a mathematical measure of intrinsic causal structure (Tononi, 2004; Oizumi et al., 2014; Albantakis et al., 2023). Predictive processing and active inference modeled perception and action as hierarchical inference and minimization of prediction error or variational free energy (Friston, 2010; Clark, 2013; Parr et al., 2022). Recurrent-processing theories emphasized feedback loops as necessary for conscious perception (Lamme & Roelfsema, 2000; Lamme, 2006). Yet none, taken alone, fully specifies why a reconstructed state should be felt as color, pain, odor, hunger, tension, pleasure, or fear.

MEM is polemical at precisely this point. Against dualism, it denies that phenomenal character requires a separate mental substance. Against classical functionalism, it argues that role-structure is too abstract: input-output organization alone does not specify the sensory quality through which an embodied organism encounters the world. Against panpsychism, it rejects primitive experiential properties where ordinary physical mechanisms may suffice. Against informational theories, it insists that information becomes phenomenally relevant only when realized as receptor-anchored, affectively weighted, and re-entrantly reconstructed activation in a living-like system.

The three core components of MEM are embodiment, motivated learning, and emotion. Embodiment means an organized physical body equipped with exteroceptors, interoceptors, proprioceptors, and effectors. Motivated learning means that action selection is guided not only by external reward but by deviations from viable allostatic ranges and by the need to reduce them over time (Starzyk, 2008; Starzyk et al., 2012; Keramati & Gutkin, 2014). Emotion is a regulatory control signal that alters salience, learning rate, memory consolidation, exploration, and policy selection.

The scientific background of MEM is inseparable from embodied intelligence, interoceptive neuroscience, and recurrent perception. Brooks (1991) and Pfeifer and Bongard (2006) emphasized situated organism–environment coupling and the shaping role of the body. Craig (2002, 2009), Damasio (1999), and Seth (2013) made bodily state and interoception central to feeling and selfhood. Lamme and Roelfsema (2000) showed that feedforward processing is insufficient for conscious vision and that recurrent interaction stabilizes perception. MEM combines these insights: exteroceptive and interoceptive signals are evaluated against allostatic needs, matched to associative structures, and used to reconstruct lower maps that enter secondary perception.

In this architecture, representations are consolidated traces of receptor-driven stimulation streams. These traces acquire emotional responses as the system react to states that improve or worsen its condition. They form layered, hetero-hierarchical associative structures called semblions, which bind a sensory prototype, a bodily-motivational context, and valence. New stimulation is compared with stored patterns and contexts; the best-fitting patterns dominate through winner-take-all or soft-competition. Recognition, recall, understanding, and inference are thus forms of associative matching and reconstruction. Thanks to these features, MEM can aspire to a theoretically grounded and empirically testable theory of phenomenal consciousness that identifies qualia and emotions with recurrently reconstructed sensory and interoceptive states.

The article formalizes this architecture by specifying the state variables of an embodied cognitive system: exteroceptive, interoceptive, and proprioceptive input, allostatic variables, bodily-motivational context, representational activity, associative links, affective modulation, curiosity, program selection, and regulatory cost. The model is not a ready-made algorithm for consciousness, but a formal framework identifying the variables required for the transition from receptor-driven perception to secondary perception, imagery, memory, affective anticipation, and conscious action.

The central contrast with standard reinforcement learning lies in the objective function. Classical reinforcement learning usually optimizes expected reward or externally defined cost. MEM begins with the organism: tolerated ranges define allostatic viability; deviations generate affective pressure; affect modulates plasticity, exploration, and valence; and action policies are selected to reduce future regulatory violations under uncertainty. Motivated learning is therefore an extension of reinforcement learning.

Within this framework, subjective sensory impressions, qualia, feelings, and emotions are physical states produced by receptor-driven and re-entrant activation of sensory and interoceptive maps. Secondary perception is the crucial mechanism: higher associative structures reconstruct lower-level sensory and bodily maps, generating felt or imagined content aligned with original perception and associated emotion. The felt quality is a structured, receptor-grounded, affectively modulated, and recurrently stabilized activation pattern.

The equations developed below organize MEM as a physicalist response to several philosophical problems. They address the explanatory gap by specifying a mechanism linking neural activation to phenomenal content; the hard problem by replacing ontological mystery with a testable re-entrant sensory-interoceptive process; the symbol-grounding problem by tying meaning to receptor history, bodily context, and action; and the limits of artificial intelligence by showing why disembodied pattern completion lacks allostatic and phenomenal grounding.
The formal components below show how associative mechanisms ground recognition, recall, understanding, and inference; how allostatic deviations generate affective control; how curiosity and exploration arise from mismatch under regulatory pressure; and how action policies are selected to limit future violations of viable bodily ranges. In this way, the mathematical model becomes both a scientific instrument and a philosophical argument.

### *1.1. Biological interpretation and scope of formalism*

The mathematical scheme below should be read as a bridge between biological description and implementable control theory. Its variables specify which biological signal classes must be jointly available for a system to recognize a situation, evaluate it relative to its needs, reconstruct missing or remembered content, and select an adaptive response. Unlike standard reinforcement learning, MEM begins with the organism: the body provides constraints, receptors provide exteroceptive and interoceptive signals, and affective pressure expresses deviations from viable ranges. This position extends embodied intelligence (Brooks, 1991, 2018; Pfeifer & Bongard, 2006) by adding interoceptive regulation and re-entrant perceptual reconstruction.

This biological interpretation also explains why cognition is not purely symbolic manipulation. The system may use symbols and programs, but their efficacy depends on sensorimotor and interoceptive history. A visual category is not merely a label attached to an image; it is a stable pattern capable of reactivating visual, tactile, motor, olfactory, gustatory, and bodily consequences. The term semblion denotes an associative unit binding a stored sensory prototype, bodily-motivational context, and valence tag. It is a distributed functional assembly, or engram, realized across autonomic loops as content and task require.

This scope is crucial for a physicalist account of phenomenal consciousness. In MEM the cost function formalizes regulatory pressure; phenomenal content is tied to activation or reactivation of sensory and interoceptive maps. Ordinary perception is receptor-driven and corrected by re-entry; imagery, dreaming, recall, and affective anticipation may reactivate the same lower maps primarily from the top down. The model therefore distinguishes affect's regulatory role from the sensory constitution of phenomenal content while remaining consistent with interoception, feeling, and recurrent processing (Craig, 2002, 2009; Damasio, 1999; Seth et al., 2011; Seth, 2013; Lamme & Roelfsema, 2000; Solms & Panksepp, 2012; Solms, 2021).

## 2. State Variables, Inputs, and Notational Conventions

In the present framework, the system is an embodied brain instantiated as a neural network that processes signals from sensory receptors and from interoceptors signaling the internal states of the system. It continuously exchanges ascending, descending recurrent, and lateral signals, and can respond either by modulating information exchange/storring or by activating effectors that enable it to act upon the environment. Time is treated as discrete: physical time is $t_k = k\Delta t$ for $k = 0,1,2,\ldots$. For notational simplicity, the subscript $k$ is suppressed and $t$ denotes the current discrete time step. The internal representational state of the system is described by the activity vector $h(t)$, with individual components $h_i(t)$ or, when the layered structure must be made explicit, $h_i^{(\ell)}(t)$.

To avoid ambiguity, we use a single symbol for each type of state variable. External receptor data are denoted by $x(t)$ and represent exteroceptive input relevant to the system's regulatory state, for example threat, unpleasant odors, excessive heat or cold, or other environmentally significant signals. Raw bodily input from interoceptors and proprioceptors is denoted by $y(t)$. It includes energetic resources, temperature, stress-related parameters, pain, visceral signals, muscle tension, and other signals originating from the body. The vector $n(t) \in \mathbb{R}^R$ denotes the allostatic state variables that are regulated by the system. The vector $b(t)$ denotes the integrated bodily-motivational and mnemonic context used in representational matching, program selection, and top-down reconstruction. Thus, $y(t)$ is the raw interoceptive report, whereas $b(t)$ is a higher-level, integrated context constructed from interoceptive, motivational, mnemonic, and task-related information.

The complete state used for action or program selection is denoted by

$$s_t := \big(h(t), b(t), n(t)\big).$$

The following notation is used consistently throughout the model:

· $x(t)$: exteroceptive input from external receptors.
· $y(t)$: raw interoceptive/proprioceptive input from the body.
· $n(t) = \big(n_1(t), \ldots, n_R(t)\big)^\top$: vector of allostatic or homeostatic state variables.
· $b(t)$: integrated bodily-motivational and mnemonic context derived from $y(t)$, $n(t)$, arousal, task mode, and relevant memory variables.
· $h(t)$: internal representational activity vector.
· $h_{\text{sens}}^{ff}(t)$: feedforward sensory representation derived from current receptor-driven input.
· $\hat{h}_{\text{sens}}(t)$: top-down reconstructed sensory representation produced by active semblions.
· $h_{\text{sens}}^{eff}(t)$: effective sensory representation after the fusion of feedforward input and top-down reconstruction.
· $S_m = (\mu_m, \rho_m, v_m)$: semblion $m$, consisting of a stored sensory prototype $\mu_m$, stored bodily-motivational context $\rho_m$, and valence $v_m$.
· $u_m(t)$: raw matching score of semblion $m$.
· $q_m(t)$: associative priming term produced by lateral links from other active semblions.
· $a_m(t)$: normalized activation share of semblion $m$.
· $L_{mn}(t)$: effective associative coupling between semblions.
· $\mathcal{C}(t)$: the set of candidate semblions selected by a Top-K or threshold rule.
· $W(t)$: the set of functionally dominant, winning semblions.
· $\mathcal{T}_j$: an elementary transition or image-schema operator.
· $\Pi$: a program, that is, a sequence or composition of elementary operators $\mathcal{T}_j$.
· $\mathcal{P}(t)$: the currently available known programs.
· $\pi(\Pi \mid s_t)$: the policy, a probability distribution over programs conditioned on the current state $s_t$.
· $Q(\Pi \mid s_t)$: the decision value of program $\Pi$ in state $s_t$.
· $J(\Pi \mid s_t)$: the predicted regulatory cost of program $\Pi$ in state $s_t$.

The connection weights include feedforward $W^{ff,\ell}$, lateral $W^{lat,\ell}$, feedback $W^{fb,\ell}$, interoceptive $W^{int,\ell}$, and contextual $W^{ctx,\ell}$ terms, used at a specific level of hetero-hierarchical structures. The functions $f$, $g$, $\psi$, $\sigma$, and $\varphi$ denote nonlinear transformations or information-combination operators; their specific form can be chosen according to the system implementation.

**Additional functions and parameters used in the model:**

· $f(\cdot)$: nonlinear state-update or activation function.
· $g(\cdot)$: function mapping regulatory violations onto global affect $A(t)$.
· $\psi(\cdot,\cdot)$: fusion function combining feedforward input and top-down reconstruction.
· $\varphi(\cdot,\cdot)$: function mapping perceptual mismatch and affect intensity onto curiosity $I(t)$.
· $\eta(t)$: dynamic learning rate; $\eta_0$ is the baseline learning rate.
· $\kappa$: gain of affective modulation of learning rate.
· $\kappa_v$: gain by which global affect updates semblion valence.
· $\varepsilon_{\min}$ and $\varepsilon_{\max}$: lower and upper limits of exploration probability in the $\varepsilon$-greedy rule.

· $\gamma$: temporal discount factor.
· $\theta_Q$, $\theta_{imp}$, and $\theta_J$: thresholds used for procedural-gap detection, improvement, and cost-based consolidation.
· $\mathrm{sim}(\cdot,\cdot)$: a similarity measure between two vectors, for example cosine similarity, normalized correlation, a Gaussian radial-basis function, or negative distance.

This notational convention separates four levels that were previously too easily conflated: raw receptor input $x(t), y(t)$; regulated allostatic variables $n(t)$; integrated bodily-motivational context $b(t)$; and the program/policy level, where programs $\Pi \in \mathcal{P}(t)$ are selected by a policy $\pi(\Pi \mid s_t)$.

## 3. Dynamics of Representation and Re-entry

Equation (1) formalizes the update of functional units—neurons, populations, or neuronal fields—under feedforward, lateral, top-down, interoceptive, and contextual influences. The layered notation keeps these streams distinct.
(1)

$$h_i^{(\ell)}(t+\Delta t) = f_i^{(\ell)}\left(A_i^{ff,\ell}(t) + A_i^{lat,\ell}(t) + A_i^{fb,\ell}(t) + A_i^{int,\ell}(t) + A_i^{ctx,\ell}(t) + \theta_i^{(\ell)}\right)$$

$$A_i^{ff,\ell}(t) = \sum_j W_{ij}^{ff,\ell}\, h_j^{(\ell-1)}(t) :$$

$$A_i^{lat,\ell}(t) = \sum_k W_{ik}^{lat,\ell}\, h_k^{(\ell)}(t) :$$

$$A_i^{fb,\ell}(t) = \sum_l W_{il}^{fb,\ell}\, h_l^{(\ell+1)}(t) :$$

$$A_i^{int,\ell}(t) = \sum_r W_{ir}^{int,\ell}\, y_r(t) :$$

$$A_i^{ctx,\ell}(t) = \sum_q W_{iq}^{ctx,\ell}\, b_q(t)$$

Here $h_i^{(\ell)}(t)$ is the activity of unit $i$ in layer $\ell$ at time $t$. The first term is the feedforward contribution from the lower layer $\ell-1$, the second term is the lateral associative contribution within the same layer, and the third term is the feedback or re-entry contribution from the higher layer $\ell+1$. The fourth term injects raw interoceptive/proprioceptive information $y(t)$, whereas the fifth term injects the integrated bodily-motivational and mnemonic context $b(t)$. The parameter $\theta_i^{(\ell)}$ is a threshold or baseline offset.

When the lowest representational layer is driven by receptors, one may set $h^{(0)}(t) \equiv x(t)$ for the exteroceptive component. The symbol $b(t)$ is reserved for the higher-level bodily-motivational context constructed from raw interoception, allostatic state, memory, and task context.
The update can be supplemented by a context-construction rule, written generically as

$$b(t+\Delta t) = \phi_b(b(t), y(t), n(t), \Pi_t),$$

where $\Pi_t$ is the program currently being executed or simulated. The role of this equation is to make explicit that $b(t)$ is not identical with raw interoception $y(t)$, but is an integrated contextual state used by semblions and by the policy of action.

Thus representation emerges in a hetero-hierarchical recurrent network that determines current activation.

## 4. Allostasis and Affect as a Regulatory Signal

In MEM, needs are encoded as violations of tolerated ranges for allostatic variables. Let $n_r(t)$ denote the current value of allostatic variable $r$, and let $n_r^{\min}(t)$ and $n_r^{\max}(t)$ denote its lower and upper tolerated bounds. A deviation below the lower bound and a deviation above the upper bound are represented by two one-sided violations:
(2)

$$\Delta n_r^-(t) = \max\left(0, n_r^{\min}(t) - n_r(t)\right)$$

$$\Delta n_r^+(t) = \max\left(0, n_r(t) - n_r^{\max}(t)\right), \quad r = 1, \dots, R$$

The complete regulatory-violation vector is

$$d(t) = (\Delta n_1^-(t), \dots, \Delta n_R^-(t), \Delta n_1^+(t), \dots, \Delta n_R^+(t))^\top.$$

Affect $A(t)$ is a signed, bounded regulatory signal derived from these violations, for instance:
(3)

$$A(t) = g\left(d(t)\right) = \tanh(w_A^\top d(t) + c_A).$$

If only the intensity of regulatory pressure is relevant, the model uses $|A(t)|$. If valence is relevant, the sign of $A(t)$ is preserved. Thus $|A(t)|$ modulates learning rate and exploration intensity, whereas $A(t)$ itself can update the valence of active semblions.

In Eqs. (2)–(3), global affect integrates allostatic deviations into a scalar control signal. It is not a particular emotion such as fear or joy, but a regulatory pressure indicating movement toward or away from tolerated states.

A more general asymmetric cost associated with the regulatory state is defined as
(4) $C(t) = \sum_{r=1}^{R}[\alpha_r(\Delta n_r^{-}(t))^2 + \beta_r(\Delta n_r^{+}(t))^2]$,
where $\alpha_r$ and $\beta_r$ allow lower-bound and upper-bound violations to have different biological significance. For example, too much pain may be strongly penalized, whereas a decrease of pain is not a violation at all. This convention is used consistently in the later cost and value equations.

Unmet needs, represented by large components of $d(t)$ or by a large value of $C(t)$, manifest as affective and motivational pressure. Such states prioritize actions or programs that are expected to restore the system to the tolerated range (Craig 2002; Seth 2013; Critchley & Garfinkel 2017; Keramati & Gutkin 2014). In animals, this relation can be tested through controlled manipulations of hunger, thirst, pain, temperature, and other regulatory needs together with measurements of valence and learning (Allen et al. 2017[1]; Betley et al. 2015; Leib et al. 2017).

### *4.1. Biological reading of allostatic thresholds and affective saturation*

The allostatic equations do not imply a single fixed equilibrium point. Biological regulation is asymmetric and context-dependent: temperature, glucose, hydration, tissue damage, blood gases, pain, fatigue, and muscular tension have tolerated ranges that may shift with posture, exertion, circadian phase, development, task demands, or social context. One-sided violations are therefore appropriate, especially for variables such as pain or suffocation.

Affect is introduced as a regulatory signal that compresses many bodily deviations into a control variable capable of modulating attention, learning, exploration, and program selection. The bounded nonlinear form is biologically motivated because neuromodulatory systems, neuronal firing, receptor sensitivity, autonomic response, and subjective intensity saturate rather than increase without limit.

Animal studies support this interpretation. Thirst- and hunger-related neural populations carry negative valence and drive learning aimed at terminating the need state (Allen et al., 2017; Betley et al., 2015; Leib et al., 2017). Computationally, this justifies treating need deviation as a motivational and prioritizing signal. MEM extends homeostatic reinforcement learning by binding regulatory pressure to semblion selection, re-entrant reconstruction, and secondary perception (Keramati & Gutkin, 2014; Galus 2026).

## 5. Semblions and Associative Matching

A semblion $S_m$ is treated here as a triple consisting of a perceptual pattern $\mu_m$ in feature space, a contextually and bodily-motivational pattern $\rho_m$, and a valence value $v_m$ representing long-term evaluative significance.
(5) $S_m = (\mu_m, \rho_m, v_m)$.
We ask how well the current configuration of sensory stimulation fits the patterns consolidated in semblions. Let $u_m(t)$ denote the degree of fit. Let $h_{\text{sens}}^{ff}(t)$ be the current feedforward sensory vector and let $b(t)$ be the integrated bodily-motivational and mnemonic context. Then:
(6)

$$u_m(t) = \alpha_s \text{sim}\left(h_{\text{sens}}^{ff}(t), \mu_m\right) + \beta_b \text{sim}(b(t), \rho_m) + \xi_v v_m$$

Here $u_m(t)$ – the outcome of matching the current feedforward sensory stimulation $h_{\text{sens}}^{ff}(t)$ and the integrated context $b(t)$ to the $m$-th semblion pattern (i.e., a measure of how well the current state corresponds to the stored perceptual pattern $\mu_m$ and the context $\rho_m$), considering the value $v_m$. It is a scalar matching score expressing the degree of correspondence between semblion $m$ and the current stimulation and contextual state.

[1] This study best illustrates the thesis that a state of homeostatic or allostatic need is not cognitively neutral but rather possesses its own negative valence and acts as an aversive motivational drive that impels the organism to act. The authors demonstrated that thirst-related neurons in the median preoptic nucleus (MnPO) are activated by dehydration, and that this activation is aversive in nature; mice learn to perform actions to terminate this activation. This strongly reinforces the central tenet of the MEM model—that a deviation from allostasis generates an affective signal which guides learning and action selection.

$h_{\text{sens}}^{ff}(t)$ – the vector of current feedforward sensory stimulation (the post-feedforward sensory map/features). This vector constitutes a single, integrated (multimodal) sensory representation at time $t$ (e.g., a vector that already incorporates vision, audition, touch, etc.).
$\mu_m$ – the sensory vector stored as the pattern encoded in semblion $m$.
$\rho_m$ – the bodily-motivational context vector associated with semblion $m$.
$v_m$ – the valence / evaluative significance associated with semblion $m$.
$b(t)$ – the integrated bodily-motivational and mnemonic context; it is not raw interoception but a context vector derived from $y(t)$, $n(t)$, arousal, task mode, and relevant memory variables.

Thus, defined memory units indicate that most mental operations in MEM are associative in character: the current state $h_{sens}$(t) is compared through similarity function sim(·,·) with stored patterns (i.e. matching of sensory maps to $\mu_m$), and the current context b(t) is compared with the stored context $\rho_m$. The resulting match, weighted by coefficients such as $\alpha,\ \beta,\ \xi$, determines the activation $u_m(t)$. In this sense, Eq. (6) is a formal statement of associative recognition, the mechanism by which the system selects the patterns/semblions best fitting the current data. What the system "recognizes" is not the raw stimulus, but the best-fitting memory pattern, anchored in patterns of past perception and in the organism's current state.

In Eq. (6), sim(·,·) is a similarity measure. The vector $\rho_m$ represents the bodily-motivational (interoceptive and situational) context in which a given semblion $S_{\text{m}}$ was encoded and for which it is most adequate. We choose the similarity measure so that it works both for two-dimensional maps (receptor arrays) and for feature vectors. The term $\text{sim}(b(t), \rho_m)$ therefore implements contextual gating: even with similar sensory content, activation is greater when the current state of the organism matches the state in which the pattern was learned.

Equation (6) formalizes associative recognition: current stimulation and context are matched to stored semblion patterns, and the best-fitting patterns are selected. Activation is treated as a continuous value, compatible with neuronal or population activity, receptor intensity, time-averaged membrane potential, or normalized activation in artificial systems, while abstracting from individual spikes (Ghosh-Dastidar & Adeli 2009; Caporale & Dan 2008).

This associative account can be supplemented by the cautiously formulated NeuroElectroDynamics (NED) hypothesis of Aur and Jog. On this view, a semblion may be implemented not only as an abstract pattern but also through charge distributions, synaptic-channel conformations, and local conduction histories in dendrites and axons.

Earlier stimulation may leave molecular and electrodynamic traces that bias later configurations toward complementary patterns (Aur & Jog 2010; Aur 2025a; 2025b). In MEM, this is a candidate micro-interpretation of matching, valence, and associative priming: excitation both activates a stored pattern and "reads" a previously modified substrate.

At the same time, NED remains only a possible biophysical interpretation. The present formalism treats symbols such as p(t), b(t), and n(t) at a mesoscopic level appropriate to the MEM architecture (Galus & Starzyk 2020).

For purposes of modeling the system and calculating its response to such activation, let $P$ denote a vector of real values. Let $vec(P)$ be an operator that reduces the dimensionality of a sensory map by flattening it into a d-dimensional real vector. Here $d$ denotes the size of the map, that is, the number of topographically organized receptor activations.

Biologically, this flattening corresponds to storing typical topographic receptor/sensory maps, for example, the photoreceptor array of the retina (retinotopic map), maps in LGN/V1 (retinotopic), somatosensory maps such as hand/finger maps in S1, or olfactory maps (glomeruli as pixels of the olfactory epithelium). After the extraction of salient features and the compression of information in higher semblion layers, the same state can be represented as a feature vector $Y$. $Y$ serves as the pattern against which sensory configurations are compared. Its components may include, for example, the activity of selected neurons in higher layers, averaged population activities, descriptors of the perceived object, or interoceptive features such as heart rate, muscle tension, pain, or arousal. The quantity $d$ then designates the number of features, or representational channels, used in that description.

The system's response is determined by a mechanism for selecting the stream of stimulation, and that selection is determined by the similarity (complementarity) of activation configurations $X$ to consolidated patterns $Y$. The similarity function may take, among others, the form of: (a) cosine similarity without mean subtraction

$$sim_{\cos}(X,Y) = \frac{X^T Y}{\| X \| \| Y \|}$$

or (b) normalized correlation with offset removal, used when maps have a background or different baseline levels

$$\tilde{x} = X - \bar{X}, \quad \tilde{y} = Y - \bar{Y}, \quad sim_{corr}(X,Y) = \frac{\tilde{x}^{\mathsf{T}} \tilde{y}}{\|\tilde{x}\| \|\tilde{y}\|}$$

In both cases the result lies in [−1, 1] and corresponds to associative matching.
The $u_m(t)$ matches alone do not yet determine which representation is to dominate in subsequent processing and become "conscious." According to MEM, selection resulting from competition among semblions is necessary.

In the MEM model, competition among active semblions is formalized by selection operators. The raw matching score $u_m(t)$, defined in Eq. (6), integrates feedforward sensory evidence, bodily-motivational context, and stored valence. Thus, the competition is not driven by bottom-up stimulation alone, but by the degree to which each semblion matches the current sensory and contextual state.

(7) $$m^*(t) = \underset{m \in \{1,\dots,M\}}{\operatorname{argmax}} \; u_m(t)$$

Equation (7) defines the hard winner-take-all case: $m^*(t)$ is the index of the semblion with the highest matching score. Under ambiguous stimulation, however, several representations may remain plausible. The TopK operator therefore selects a candidate set containing the $K$ highest-scoring semblions:

(8) $$C(t) = \mathrm{TopK}_K(\{u_m(t)\}_{m=1}^{M})$$

Here, $K$ is a fixed or implementation-dependent selection parameter specifying the maximum number of semblions retained for subsequent competition. The set $C(t)$ should be distinguished from the final winning set $W(t)$, defined later by the dominance threshold in Eq. (12).
A
soft-competition variant is then applied within the candidate set $C(t)$. Lateral associative interactions modify the raw matching score before normalization:

(9) $$q_m(t) = \sum_{n \in C(t)\setminus\{m\}} L_{mn}(t)\, [u_n(t)]_+$$

$$[z]_+ := \max(0, z)$$

$$\tilde{u}_m(t) = u_m(t) + \lambda_L q_m(t)$$

$$a_m(t) = \frac{\exp(\tilde{u}_m(t)/\tau_c)}{\sum_{j \in C(t)} \exp\left(\tilde{u}_j(t)/\tau_c\right)}, \qquad m \in C(t)$$

$$a_m(t) = 0, \qquad m \notin C(t)$$

Here, $q_m(t)$ denotes associative priming of semblion $m$, $L_{mn}(t)$ is the effective associative coupling between semblions $m$ and $n$, $\tilde{u}_m(t)$ is the association-modulated matching score, and $a_m(t)$ is the normalized activation share. The coefficient $\lambda_L$ controls the strength of lateral associative priming, whereas $\tau_c > 0$ is the softmax temperature. Lower values of $\tau_c$ produce sharper, WTA-like competition, whereas higher values allow several candidate representations to remain simultaneously active.

The distinction between $C(t)$ and $W(t)$ is important. $C(t)$ contains representations admitted to the competitive stage, whereas $W(t)$ contains those that subsequently reach sufficient dominance to influence top-down reconstruction and secondary perception.

Many semblions may remain partially active, but only those reaching sufficient dominance influence top-down reconstruction and effective perception. In the hard WTA version, the winner $m^*(t)$ is the semblion that maximizes $u_{\mathrm{m}}$. In implementation, one may keep the top-k winners.

MEM may select winners from the association-modulated score $\tilde{u}_m(t)$. In this way, a semblion that is only moderately supported by bottom-up stimuli may still enter the winning coalition if it is strongly primed by other currently active semblions through previously consolidated associations $L_{mn}$. This provides a formal mechanism for lateral or multimodal perceptual completion. For example, activation of a visual semblion corresponding to apple may increase the activation of associated olfactory, gustatory, motor, or interoceptive semblions, thereby enriching the percept beyond the information directly available in the current sensory stream.

Let $L_{mn}(t)$ denote the effective intermodal, motor, and interoceptive, associative coupling between semblions $m$ and $n$. This quantity should be understood as a macroscopic parameter summarizing repeated co-activation, temporal coincidence, stabilization of synaptic microcircuits, dendritic-spine remodeling, astrocytic modulation, and pruning-like selection of effective pathways. In this sense, $L_{mn}(t)$ expresses how strongly the activation of semblion $n$ facilitates the activation or reconstruction of semblion $m$.

(10) $$L_{mn}(t + \Delta t) = (1 - \gamma_L)L_{mn}(t) + \eta_L(t)a_m(t)a_n(t).$$

(11) $$\eta_L(t) = \eta_{L0}(1 + \kappa_L|A(t)|).$$

Correspondingly, $\eta_{L0}$, $\kappa_L$, $\gamma_L$ and affective terms determine, respectively, the baseline speed of link formation and the influence of affect.

Therefore, $L_{mn}(t)$ encompasses associations among visual, auditory, olfactory, gustatory, somatosensory, motor, and interoceptive representations. It can sustain semantic, episodic, sensorimotor, and affective associations, including cross-modal ones. The update rule $L_{mn}(t)$ therefore describes the strengthening or weakening of the associative disposition between assemblies of semblions.

The increase of $L_{mn}(t)$ results from the repeated co-activation of dominant semblions and is modulated by affective salience, while a decrease reflects the decay of ineffective associations. Consequently, those associations that are frequently co-activated and behaviorally significant are retained.

$W(t)$ - the set of "winning" semblions, selected by a threshold $\Theta$ or by the top-k rule is:

(12) $$W(t) = \{m\colon\ a_m(t) > \Theta\}.$$

where the threshold $\Theta$ specifies the minimal activation share required to influence **top-down reconstruction** and **secondary perception**. $W(t)$ formalizes the competition among perceptual stimulation described in MEM.

At this point, we should clarify the distinction between re-entry, reconstruction, and secondary perception. In this work, re-entry is a mechanism, reconstruction is a process, and secondary perception is the phenomenological outcome of that process.

The percept therefore becomes an associatively stabilized configuration: an activated visual semblion can prime other modalities through the lateral field, so an object is represented as a multimodal network consolidated by repeated co-activation.

### *5.1. Semblions, modality specificity, and the association problem*

A semblion is a distributed associative coalition that may include modality-specific subassemblies and cross-modal links while preserving visual, auditory, tactile, olfactory, motor, and interoceptive codes.

The matching rule expresses the same principle mathematically. The sensory prototype reflects prior receptor-driven encounters; the context prototype captures bodily and task state; and valence reflects long-term regulatory significance. Understanding, in this biologically grounded sense, is the selection of a memory-based pattern fitting both sensory configuration and organismic state. The same object may therefore be represented differently in hunger, satiety, fear, curiosity, pain, or fatigue.

This interpretation also explains the use of both hard and soft selection. In some situations, a single pattern dominates rapidly, approximating a winner-take-all regime. In ambiguous situations, several candidate semblions may remain partially active. The softmax formulation describes a plausible biological regime in which attention, arousal, and neuromodulatory state regulate the sharpness of competition. When the softmax temperature $\tau_c$ is low (9), competition is sharp, and behavior is decisive. When the temperature is high, multiple hypotheses remain active, allowing perception to be refined by context, memory, and further sampling. This view is consistent with decision-bound models in which neural populations accumulate evidence until a flexible threshold is crossed (Hawkins et al., 2014; Simen, 2012; Wong & Wang, 2007).

The associative-coupling term can prime other frequently co-activated semblions: seeing a lemon may reactivate visual features, sour taste, smell, salivation, grasping programs, facial reactions, and valence. The model remains compatible with spiking and spike-timing-dependent approaches while abstracting to population-level activation (Caporale & Dan, 2008; Ghosh-Dastidar & Adeli, 2009). It also accommodates the active role of the substrate in semblion processing, including Neuro-Electro-Dynamics (Aur & Jog 2010; Aur 2025a, 2025b) and mechanisms such as synaptogenesis, ephaptic coupling, dendritic-spine modulation, tripartite synapses, and epigenetic modification (Galus 2025b).

## 6. Top-down Reconstruction and Secondary Perception

If a set of winning semblions W(t) has been selected, the reconstructed sensory map is a combination of their prototypes $a_{\mathrm{m}}$.

(13) $$\hat{h}_{\mathrm{sens}}(t) = \sum_{m \in W(t)} a_m(t) \mu_m.$$

The reconstructed signal $\hat{h}_{sens}$ represents sensory content retrieved from associative memory.
Higher-level cognitive layers of active semblions impose an activation pattern on lower sensory maps, which then undergo secondary perception, interpreted within MEM as phenomenal "experience." Equation (13) thus formalizes re-entry, through which stored sensory content is reconstructed and made available to this form of perception.
In the hard WTA version, the sum reduces to the single winning semblion.

(14) $$h_{\mathrm{sens}}^{eff}(t) = (1-\lambda) h_{\mathrm{sens}}^{ff}(t) + \lambda \hat{h}_{\mathrm{sens}}(t).$$

The effective perceptual representation $h_{sens}^{eff}(t)$ is the fusion of the feedforward signal and the reconstruction, that is, the final representation available for further processing and, in MEM, for constituting phenomenal content.
Equation (14) presents the simplest weighted-average form of the fusion function. More generally, $\psi(\cdot,\cdot)$ may denote any operator integrating feedforward input and top-down reconstruction. The term $h_{\mathrm{sens}}^{ff}$ denotes the bottom-up signal arriving from receptors and early processing stages, while $\hat{h}_{\mathrm{sens}}$ denotes the top-down reconstruction imposed by memory and active semblions. The weighting parameter $\lambda$ controls the relative contribution of bottom-up and top-down information.

It determines the relative strength of feedback in relation to bottom-up signals, as well as the influence of attention, arousal, and neuromodulation. In the MEM model, it reflects the influence of affective state: at low values of $\lambda$, perception is primarily **data-driven**; at high values, it is **memory-driven**. Equation (14) can therefore cover ordinary perception, priming and context effects, imagination, and hallucination-like or illusory cases in which reconstruction increases and the bottom-up signal is weak or noisy.

In the MEM framework, Image Schemas (IS) (for example notions: CONTAINER, SOURCE–PATH–GOAL, FORCE, or BALANCE) are modeled as elementary transition operators $\mathcal{T}_j$. These operators should be distinguished from programs $\Pi$. An elementary transition operator maps the current state and an action into a subsequent state; a program is a sequence or composition of such elementary operators.

(15) $$\big(h(t+\Delta t), b(t+\Delta t), n(t+\Delta t)\big) = \mathcal{T}_j(h(t), b(t), n(t), a_t).$$

A program is defined as a finite composition of elementary operators,

$$\Pi = \big(\mathcal{T}_{j_0}, \mathcal{T}_{j_1}, \dots, \mathcal{T}_{j_T}\big),$$

and an episode $E$ is the realized or simulated trajectory of states, actions, and applied operators:

(16) $$E = \big(s_0, a_0, \mathcal{T}_{j_0}, s_1, a_1, \mathcal{T}_{j_1}, \dots, s_T\big).$$

An episode records what happened, or what was simulated as happening, when operators were applied to an initial state. It includes states, elementary actions, operators, and bodily-regulatory and affective consequences. A program is a reusable procedural structure abstracted from episodes or generated by counterfactual simulation and later selected by the policy. Thus, episodes are descriptive and retrospective, whereas programs are procedural and prospective.

In the MEM model, the winning semblion $m$ may also reconstruct semblions $n$ that are modally linked or contextually associated, weighted by normalized associative couplings $\tilde{L}_{mn}(t)$. Consequently, re-entry not only reproduces the matched pattern but also supplements the percept with information that frequently co-occurs in experience.

This is a mechanistic explanation of cross-modal completion in perception and imagination. Visual recognition of a lemon may reactivate gustatory, olfactory, somatosensory, motor, and interoceptive prototypes, enriching reconstruction with anticipated sourness, scent, salivation, grasping tendencies, and missing fragments of the original sensory representation.

6.1. *Secondary perception*

The reconstruction equations (13) are central to the physicalist interpretation of MEM. Re-entry is not merely computational correction but the reactivation of lower sensory and interoceptive maps. We claim that when reactivation is coherent and functionally dominant, the system undergoes secondary perception: the pattern is experienced as felt or imagined content. Feedforward activity supplies evidence, lateral links stabilize associations, feedback drives reconstruction, and the fused representation becomes available for cognition and action.

MEM preserves a distinction often lost in predictive accounts. Prediction error may guide learning, but it does not explain phenomenal character. MEM locates phenomenal content in the reactivation of sensory and interoceptive fields continuous with receptor-driven perception. Imagery, dreams, anticipatory emotion, and memory are therefore forms of re-entrant reconstruction.

The fusion parameter has a clear biological meaning. Attention, arousal, sensory noise, neuromodulation, task demands, and affective state can shift the balance between reliable bottom-up input and memory-based reconstruction. The same framework spans direct perception, perceptual completion, emotionally biased interpretation, imagery, dreaming, and hallucination-like states.

The claim is testable. Perturbing feedback routes to early sensory or interoceptive maps should affect vividness and qualitative character of imagery, recall, and conscious perception more than fast unconscious feedforward discrimination. Strengthening re-entrant coupling should increase vividness, stability, or affective coloring. MEM does not deny higher-order report, global availability, or information integration; it argues they are insufficient without embodied sensory-interoceptive reconstruction.

## 7. Motivated Learning: Modulation of Learning Rate and Valence

In MEM, affect $A(t)$ is a control signal. It can increase the effective learning rate $\eta(t)$ in critical situations and update the valence $v_{\mathrm{m}}$ associated with a semblion. This corresponds to the intuition that events important for survival are remembered more strongly and more rapidly.

Let $\eta(t)$ denote the instantaneous (dynamic) learning rate/plasticity at time $t$, and $\eta_0$ the baseline learning rate when affect is zero. $A(t)$ is the global affect integrating the pressure of needs, and $|A(t)|$ is affect intensity regardless of sign. A crucial role is played by the gain coefficient $\kappa$, that is, the sensitivity of the learning rate to affect.

(17)
$$\eta(t) = \eta_0(1 + \kappa|A(t)|).$$

The coefficient $\kappa$ has major significance in the MEM model. It specifies how strongly the affect modulates the learning rate. The larger κ is, the more strongly the system "learns faster" under high regulatory pressure (large $|A(t)|$), which corresponds to biological effects of neuromodulation that increase plasticity and prioritize encoding.

If $\kappa = 0$, there is no modulation and $\eta(t) = \eta_0$, that is, emotionally neutral learning. If $\kappa > 0$, a stronger affect leads to greater plasticity and hence faster adaptation. In principle $\kappa < 0$ is mathematically possible, but within MEM a sensible regulatory interpretation normally assumes $\kappa \geq 0$.

(18)
$$v_m(t + \Delta t) = \mathrm{clip}_{[v_{\min}, v_{\max}]}\big((1 - \lambda_v)v_m(t) + \eta(t)\kappa_v A(t) a_m(t)\big).$$

Valence $v_m$ is associated with the semblion $S_m$: it codes whether the pattern or association has historically been beneficial, neutral, or detrimental for regulatory stability. The update is now tied to the activation share $a_m(t)$, so affect modifies primarily those semblions that are active. The coefficient $\lambda_v \in [0,1]$ allows slow decay or relaxation of old valence tags, while $\mathrm{clip}_{[v_{\min}, v_{\max}]}$ prevents unbounded drift. The signed affect $A(t)$ determines whether the active representation is tagged positively or negatively; its magnitude is already incorporated into the dynamic learning rate $\eta(t)$ from Eq. (17).

## 8. Action Policy and Objective Function

In MEM, the policy selects programs, not isolated actions. An elementary action $a_t$ may be a single step inside a program, but the decision variable at the regulatory level is a program $\Pi$, understood as a sequence or composition of elementary transition operators $\mathcal{T}_j$. The policy is therefore written as a probability distribution over programs conditioned on the full state $s_t = \big(h(t), b(t), n(t)\big)$.

(19) $$\Pi_t \sim \pi(\cdot \mid s_t), \qquad s_t = \big(h(t), b(t), n(t)\big).$$

If an elementary action is needed at the effector level, it is the first executable act prescribed by the selected program:

(19a) $$a_t = \mathrm{first}(\Pi_t).$$

The policy is learned to minimize the expected discounted regulatory cost generated by the programs it selects:

(20) $$\pi^* = \underset{\pi}{\mathrm{argmin}}\,\mathbb{E}_{\Pi\sim\pi}\left[\sum_{\tau=0}^{\infty} \gamma^\tau\, C(t+\tau\Delta t; \Pi) \mid s_t\right]$$

Here the expectation is taken over trajectories (episodes $\mathbb{E}_{\Pi\sim\pi}$) generated by the selected programs and, when the environment is stochastic, over possible state transitions. The instantaneous regulatory cost is the asymmetric violation cost introduced in Eq. (4):

(21) $$C(t; \Pi) = \sum_{r=1}^{R}\left[\alpha_r(\Delta n_r^-(t;\Pi))^2 + \beta_r(\Delta n_r^+(t;\Pi))^2\right].$$

The predicted cost of a program over a finite planning horizon $T$ is

(21a) $$J(\Pi \mid s_t) = \mathbb{E}\left[\sum_{\tau=0}^{T} \gamma^\tau\, C(t+\tau\Delta t; \Pi) \mid s_t\right]$$

The finite-horizon form is used in program comparison and procedural-gap detection, because real organisms and artificial systems operate under time and computational constraints.

Under strong regulatory pressure, instead of searching over all programs, the system may restrict the search to programs associated with the most urgent allostatic violation. Let

$$r^*(t) = \underset{r}{\mathrm{argmax}}\max\left(\big(\Delta n_r^-(t)\big)^2, \big(\Delta n_r^+(t)\big)^2\right)$$

identify the currently dominant regulatory threat, and let $\mathcal{P}_{r^*}(t) \subseteq \mathcal{P}(t)$ be the subset of programs relevant to that threat. Then the urgent-response rule is

(22) $$\Pi_t^* = \arg\min_{\Pi\in\mathcal{P}_{r^*}(t)} J(\Pi \mid s_t).$$

The immediate urgency cost can be written as

(23) $$J_{\max}(t) = \max_r\left(\big(\Delta n_r^-(t)\big)^2, \big(\Delta n_r^+(t)\big)^2\right).$$

Equations (19)–(23) distinguish clearly between the policy $\pi$, which is a probability distribution over programs, and the program repertoire $\mathcal{P}(t)$, which is the set of known programs available for selection.

## 9. Reconstruction Error, Curiosity, and Exploration Mode

The difference between the feedforward signal and top-down reconstruction may serve as a measure of novelty and/or uncertainty:

(24) $$r(t) = \left\| h_{\mathrm{sens}}^{ff}(t) - \hat{h}_{\mathrm{sens}}(t) \right\|^2.$$

The quantity $r(t)$ is the perceptual-deviation error (primary/direct or secondary perception): it measures the discrepancy between current sensory stimulation $h_{\mathrm{sens}}^{ff}(t)$ and the patterns $\hat{h}_{\mathrm{sens}}(t)$ consolidated by the selected semblions or image schemas. A high value of $r(t)$ indicates novelty, uncertainty, or a stimulus that is poorly explained by its current schemas, which in MEM becomes an impulse to exploration and representational updating.

The curiosity function $I(t)$ combines the mismatch between observation and the known image schema, r(t), with the affect intensity $|A(t)|$.

(25) $$I(t) = \varphi(r(t), |A(t)|).$$

The quantity $I(t)$, normalized between 0 and 1, combines perceptual-deviation error as novelty or schema mismatch with affect intensity as regulatory pressure. High values indicate situations that are both informationally difficult and organismically significant.

Curiosity function regulates the exploration parameter $\varepsilon(t)$, for example in $\varepsilon$-greedyalgorithms or exploration/exploitation policies.

(26) $$\varepsilon(t) = \varepsilon_{\min} + (\varepsilon_{\max} - \varepsilon_{\min})I(t).$$

The variable $\varepsilon(t)$ is the adaptive exploration probability in an $\varepsilon$-greedy scheme, increasing with curiosity I(t). For I(t) = 0, exploration equals $\varepsilon_{\min}$; for I(t) = 1, it reaches $\varepsilon_{\max}$. This mechanism corresponds to switching between exploitation of existing schemas and attempts at new actions. Similar mechanisms had previously been used in reinforcement learning (Sutton & Barto 1998).

The perceptual-deviation error $r(t)$ from Eq. (24) can be interpreted as the "price of mismatch" between feedforward signal $h_{sens}^{ff}(t)$ and memory-based reconstruction $\hat{h}_{sens}(t)$: the worse $\hat{h}_{sens}(t)$ explains $h_{sens}^{ff}(t)$, the larger $r(t)$ becomes, which, together with regulatory pressure $A(t)$, raises curiosity $I(t)$ in Eq. (25) and consequently increases the level of exploration ε(t) in Eq. (26).

*9.1. Curiosity, exploration, and uncertainty under regulatory pressure*

Curiosity in MEM is not free-floating novelty seeking. It arises when perceptual mismatch matters for regulation. A mismatch may be harmless, but under hunger, danger, pain, thirst, social threat, or absence of a usable program, it becomes urgent.

Unlike the fixed exploration parameter often used in introductory reinforcement-learning algorithms (Sutton & Barto, 1998), MEM makes exploration state-dependent. Low mismatch and low affect favor exploitation of familiar programs; high mismatch combined with regulatory pressure raises exploratory probability, constrained by time, energy, and learned schemas. Exploration is thus a regulated policy shift
T
he same formalism offers a mechanistic explanation why a novel but irrelevant stimulus may be ignored in safety, yet a minor cue is investigated under thirst, threat, or uncertainty. It is consistent with motivated-learning approaches in which autonomous systems generate and revise goals (Starzyk, 2008; Starzyk et al., 2010; Starzyk, 2011; Starzyk et al., 2012, 2013).

## 10. Procedural Gap, Creativity, and New Ways of Reacting

Creativity in MEM is formalized as a procedural gap: this appears when in the current state, the system lacks a known program of sufficiently high decision value. Value is defined as negative predicted regulatory cost; therefore, higher value means a better expected reduction of future allostatic violations.

(27) $$Q(\Pi \mid s_t) := -J(\Pi \mid s_t) = -\mathbb{E}[\sum_{\tau=0}^{T} \gamma^{\tau}\, C(t+\tau\Delta t;\Pi) \mid s_t]$$

With this convention, larger $Q$ and smaller $J$ is better. This removes the ambiguity between value and cost.
The procedural-gap indicator is defined as

(28) $$K(t) = 1 \quad \Leftrightarrow \quad \max_{\Pi\in\mathcal{P}(t)} Q(\Pi \mid s_t) < \theta_Q < 0.$$

Thus $K(t) = 1$ marks the moment when the currently available repertoire $\mathcal{P}(t)$ contains no program whose predicted value exceeds the sufficiency threshold $\theta_Q$. This triggers a candidate-generation mode: new transition operators $\mathcal{T}_j$ or new compositions of operators are proposed, producing candidate programs $\Pi \in \mathcal{P}_{cand}(t)$. The policy $\pi$ is a distribution over programs; the repertoire $\mathcal{P}(t)$ is the set of known programs; and $\mathcal{P}_{cand}(t)$ is the set of newly generated candidate programs.

A newly generated program is selected by maximizing value, equivalently by minimizing predicted regulatory cost:
(29)

$$\Pi_{new}(t) = \arg\max_{\Pi\in\mathcal{P}_{cand}(t)} Q(\Pi \mid s_t) = -\arg\min_{\Pi\in\mathcal{P}_{cand}(t)} J(\Pi \mid s_t)$$

If the candidate is retained, the program repertoire is updated as

(30) $$\mathcal{P}(t+\Delta t) = \mathcal{P}(t) \cup \{\Pi_{new}(t)\}.$$

The planning horizon $T$ is determined as a compromise between (i) the time window $B$ over which actions can influence the reduction of tolerance-bound violations and (ii) the computational limitations of simulation. In practice, $T$ may be fixed or adaptive: shortened in states of high regulatory pressure and extended in stable states to allow prospective simulation and the selection of more complex programs. If $M$ candidate programs must be evaluated within time budget $B$, the evaluation time per candidate is

(31) $$T_{eval} = \frac{B}{M}.$$

Accordingly, the number of simulated steps per candidate can be approximated by

(32) $$T \approx \left\lfloor \frac{T_{eval}}{\Delta t} \right\rfloor.$$

Let $\Pi_{best}(t) = \arg\max_{\Pi\in\mathcal{P}(t)} Q(\Pi \mid s_t)$

be the best-known program in the current repertoire.
Improvement is now defined unambiguously as a positive difference in decision value:

(33) $$\Delta Q(t; \Pi_{new}, \Pi_{best}) = Q(\Pi_{new} \mid s_t) - Q(\Pi_{best} \mid s_t)$$

A positive $\Delta Q$ means that the new program is predicted to reduce regulatory cost more effectively than the best-known program. Consolidation of a newly generated program is therefore written as

(34) $$\mathcal{P}(t + \Delta t) = \mathcal{P}(t) \cup \{\Pi_{new}(t)\} \quad \text{if} \quad \Delta Q(t; \Pi_{new}, \Pi_{best}) \geq \theta_{imp} > 0$$
$$\mathcal{P}(t + \Delta t) = \mathcal{P}(t) \quad \text{otherwise}$$

An equivalent cost-threshold formulation is

(34a)
$$\mathcal{P}(t + \Delta t) = \mathcal{P}(t) \cup \{\Pi_{new}(t)\} \quad \text{if} \quad J(\Pi_{new} \mid s_t) \leq \theta_J.$$

Equations (33)-(34a) eliminate a possible sign ambiguity by defining improvement as a positive increase in value, whereas sufficiently low predicted cost is expressed separately by the inequality involving the predicted cost threshold. Creativity is therefore productive only when a new candidate program becomes a stabilized competence: it must either improve value relative to the best-known program or satisfy an absolute regulatory-cost threshold $\Delta Q(\Pi_{new} \mid s_t) \leq \theta_J$

### *10.1. Procedural gaps, creative programs, and biological consolidation*

The procedural-gap condition gives MEM a formal account of creativity. A new program is generated because the existing repertoire lacks a program with sufficient predicted value for the current state. Creativity is therefore an adaptive response to insufficiency, constrained by regulatory pressure and by the need to balance exploration with control.

The distinction between transition operator and program is biologically important. A transition operator is a learned or simulated local transformation—grasping, avoiding, approaching, vocalizing, shifting attention, testing a surface, or changing posture. A program is a structured sequence of such transformations, informed by episodic trajectories, procedural routines, and prospective simulation. Mathematically, the program is evaluated by predicted regulatory cost, not verbal plausibility.

Consolidation closes the loop from exploratory creativity to competence. A candidate is retained only if it improves value or satisfies a low-cost threshold, preventing learning from accidental noise. A successful solution should strengthen participating semblions, update valence, and increase reactivation in similar contexts. Thus, what is felt, remembered, imagined, and planned becomes significant insofar as it can alter future allostatic trajectories.

### *10.2. Testability, falsification, and limits of the model*

The model is empirically accountable. MEM would be supported if interoceptive or allostatic changes altered choices, matching, valence tagging, imagery vividness, and exploration–exploitation balance as predicted. It would also be supported if feedback perturbations changed secondary perception and phenomenal report while sparing fast unconscious feedforward detection. It would be weakened if conscious imagery and affective recall proved independent of sensory or interoceptive reactivation, or if allostatic manipulations failed to affect learning, valence, and policy selection after reward variables were controlled.

The model is limited in scope. It offers a mesoscopic formal framework: sensory maps, context vectors, semblions, associative couplings, regulatory violations, affective gain, reconstruction, curiosity, and program selection.
MEM proposes a candidate mechanism for narrowing the explanatory gap: receptor-driven and re-entrant activation of sensory and interoceptive maps supplies phenomenal content, while allostatic regulation supplies affective significance. Its advantage is that it specifies what must be measured, perturbed, and simulated; its risk is that future evidence may reveal conscious contents not dependent on this reconstruction mechanism.

## 11. Summary and Perspectives for Further Research

MEM is inspired by biological processes observed in natural brains. Embodiment provides the physical substrate necessary for the generation of qualia through recurrent reconstruction of sensory and interoceptive representations. Interoceptive signals indicate allostatic deviations and generate feelings and emotions; competition among bottom-up stimulations, categorization, generalization, and associations across semblion layers form object representations and a world model through motivated learning. Recursive processes and secondary perception generate memories, dreams, hallucinations, and visualization of imagined objects, enabling planning and conscious decision-making.

Many of these processes are supported by neurological experiments and appear in other models of mind. MEM argues that only their integration within one embodied, affectively regulated, recurrent architecture provides a formal physicalist account for the organization of the human and animal psyche.

The mathematical model captures motivated learning as a regulatory process operating in an extended state space S × B × N × A: cognitive/sensory representations, bodily-motivational contexts, allostatic variables, and affective control signals. Its nonlinear dynamics depend on perceptual, associative, and executive operators modulated by global affect, which is itself a function of regulatory deviations from tolerance thresholds.

Learning is therefore not classical reward maximization but minimization of a cost functional dominated by discounted deviations of regulatory variables from preferred ranges.

Motivated learning is formulated as the problem of minimizing a regulatory-cost function that depends on deviations of allostatic variables from tolerance thresholds, whether symmetric or asymmetric. Global affect $A(t)$, being a nonlinear function of the vector of regulatory violations $d(t)$, functions as a modulatory signal affecting the learning rate η, the updating of representational valence, and the switching between exploration and exploitation. In the MEM model, affect is one of the central regulators. It is described by equations (2) – (4), while the effects of affect are captured in equations (17) and (18). Deviations from allostasis/homeostasis generate a regulatory-violation vector $d(t)$, which subsequently gives rise to global affect $A(t)$. This constitutes a global valence signal that integrates the multidimensional regulatory state into a single scalar pressure for action.

The action policy $\pi(\Pi \mid s_t)$ minimizes the expected, discounted cost of future deviations over a planning horizon $T$. A procedural gap is defined as the condition in which no known program $\Pi \in \mathcal{P}(t)$ exceeds a decision-value threshold. This initiates the generation of new elementary transition operators $\mathcal{T}_j$ or new program compositions $\Pi$ through counterfactual simulation and selection that minimizes predicted regulatory cost.

The formalism introduces feedback between representational state and plasticity: learning rate and hypothesis selection depend on the system's regulatory state. The system therefore modifies not only its representations but also its adaptation rules, so stability depends on tolerance thresholds, modulation coefficients, and planning horizon.

The model is a formal skeleton integrating perception, association, homeostatic regulation, and prospective planning into a unified optimization structure. It is not an algorithm for generating consciousness, but a set of necessary conditions for adaptive, motivated learning in a dynamic environment. Because mathematical description allows multiple realization, it leaves open the possibility of artificial structures capable of feeling and empathy.

Open problems include stability analysis, identification of conditions under which the model reduces to optimal-control or motivated-learning schemes, and clarification of how a regulatory objective function differs from a reward function. The MEM model indicates that a percept is not a neutral image of the world. The same object may possess varying activation strength and a different perceptual status depending on hunger, pain, anxiety, arousal, task state, and learning history (pre-existing patterns). The thesis emerges that the neuronal representation of a percept is a hypothesis regarding the state of the world and the organism. Does MEM explain the nature of **C**onsciousness? The parameter "**C**" appears nowhere in the equations nor in the text of the article. Nevertheless, the presented mathematical description of the formation of perceptual representations aligns with the thesis that conscious perception is essentially the recognition of an object that elicits complex emotional responses, stimulates learned reactions, and updates the model of the surrounding reality

Further work should ground the generation of new transition operators under regulatory constraints, analyze the time scales separating perception, plasticity, and homeostatic change, and address parameter identifiability. Numerical simulations and empirical data will determine whether formalism has the properties necessary and sufficient for stable, motivated learning and conscious action in dynamic environments.

The model may then help interpret longstanding philosophical and psychological problems with greater precision: free will, the explanatory gap, mind–brain identity theory, and whether artificial intelligence could ever attain a depth of mental phenomena comparable to that of the human mind.

## Bibliography

Albantakis, L., *et al.* (2023). Integrated information theory (IIT) 4.0: Formulating the properties of phenomenal existence in physical terms. *PLOS Computational Biology*, 19(10), e1011465.

Allen, W. E., *et al.* (2017). Thirst-associated preoptic neurons encode an aversive motivational drive. *Science,* 357(6356).

Ashby, W. R. (1952). *Design for a Brain*. Chapman & Hall.

Aur, D. (2025a) When Matter Thinks: The Physics of Experience in Electrodynamic Intelligence Systems. *Zendo* DOI: 10.5281/zenodo.17087828.

Aur, D., (2025b) The Second Wave: From Neuroelectrodynamics to Electrodynamic Intelligence. *TechRxiv.* DOI: 10.36227/techrxiv.175623132.24668882/v1

Aur, D., & Jog, M. S. (2010). *Neuroelectrodynamics - Understanding The Brain Language.* IOS Press.

Baars, B. J. (1988). *A Cognitive Theory of Consciousness.* Cambridge University Press.

Betley, J. N., *et al.* (2015). Neurons for hunger and thirst transmit a negative-valence teaching signal. *Nature*, 521(7551), 180-185. https://doi.org/10.1038/nature14416

Brooks, R. A. (1991). Intelligence without representation. *Artificial Intelligence*, 47(1-3), 139-159.

Brooks, R. A. (2018). *Intelligence without reason. In The Artificial Life Route to Artificial Intelligence* (pp. 25-81). Routledge.

Caporale, N., & Dan, Y. (2008). Spike timing-dependent plasticity: A Hebbian learning rule. *Annual Review of Neuroscience*, 31(1), 25-46. https://doi.org/10.1146/annurev.neuro.31.060407.125639

Chalmers, D. J. Facing up to the problem of consciousness. *Journal of Consciousness Studies*, 2(3), 200-219, (1995).

Chalmers, D. J. (1996). *The Conscious Mind: In Search of a Fundamental Theory*. Oxford University Press.

Clark, A. (2013). Whatever next? Predictive brains, situated agents, and the future of cognitive science. *Behavioral and Brain Sciences*, 36(3), 181-204.

Craig, A. D. (2002). How do you feel? Interoception: The sense of the physiological condition of the body. *Nature Reviews Neuroscience*, 3, 655-666. https://doi.org/10.1038/nrn894

Craig, A. D. (2009). How do you feel now? The anterior insula and human awareness. *Nature Reviews Neuroscience*, 10(1), 59-70. https://doi.org/10.1038/nrn2555

Critchley, H. D., & Garfinkel, S. N. (2017). Interoception and emotion. *Current Opinion in Psychology*, 17, 7-14.

Damasio, A. R. (1999). *The Feeling of What Happens: Body and Emotion in the Making of Consciousness*. A Harvest Book.

Dehaene, S., & Changeux, J.-P. (2011). Experimental and theoretical approaches to conscious processing. *Neuron*, 70(2), 200-227.

Fechner, G. T. (1966). *Elements of Psychophysics* (H. E. Adler, Trans.). Holt, Rinehart and Winston. (Original work published 1860)

Feigl, H. (1958). The 'mental' and the 'physical'. In H. Feigl, M. Scriven, & G. Maxwell (Eds.), Concepts, Theories, and the Mind-Body Problem. University of Minnesota Press.

Friston, K. (2010). The free-energy principle: A unified brain theory? Nature Reviews Neuroscience, 11, 127-138.

Galus, W. L. (2018). Semblions of Words: The Language of Natural and Artificial Neural Networks. *Qeios*, 1ATS9M. https://doi.org/10.32388/1ATS9M.2

Galus, W. L. (2023a). Different aspects of consciousness explained by distinct biophysical processes. *Journal of Theoretical and Philosophical Psychology*. Advance online publication. https://doi.org/10.1037/teo0000236

Galus, W. L. (2023b). Mind-brain identity theory confirmed? *Cognitive Neurodynamics*, 17, 1467-1487. https://doi.org/10.1007/s11571-023-09992-6

Galus, W.L. (2025a) Perception as the Essence of Phenomenal Consciousness. Available at *SSRN*: http://dx.doi.org/10.2139/ssrn.5207381

Galus, W.L., (2025b) A new physicalist model of consciousness: a proposal. Available at *SSRN*: DOI: http://dx.doi.org/10.2139/ssrn.5207389

Galus, W.L (2026). The Concept of a Motivated Emotional Mind Explains Thinking Processes in Cognitive and Phenomenal Aspects. *Integr. Psych. Behav.* **60**, 32 (2026).

Galus, W. L., & Starzyk, J. A. (2020). *Reductive Model of the Conscious Mind*. IGI Global.

Ghosh-Dastidar, S., & Adeli, H. (2009). Spiking neural networks. International Journal of Neural Systems, 19(04), 295-308.

Green, D. M., & Swets, J. A. (1966). *Signal Detection Theory and Psychophysics*. Wiley.

Hawkins, G. E., *et al.* (2014). Neural dynamics implement a flexible decision bound with a fixed firing rate for choice: A model-based hypothesis. *Frontiers in Neuroscience*, 8, 318.
Keramati, M., & Gutkin, B. (2014). Homeostatic reinforcement learning for integrating reward collection and physiological stability. *eLife*, 3, e04811.
Lamme, V. A. F. (2006). Towards a true neural stance on consciousness. *Trends in Cognitive Sci.*, 10(11), 494-501.
Lamme, V. A. F., & Roelfsema, P. R. (2000). The distinct modes of vision offered by feedforward and recurrent processing. *Trends in Neurosciences*, 23(11), 571-579.
Lau, H., & Rosenthal, D. (2011). Empirical support for higher-order theories of conscious awareness. *Trends in Cognitive Sciences*, 15(8), 365-373.
Leib, D. E., *et al.* (2017). The forebrain thirst circuit drives drinking through negative reinforcement. *Neuron*, 96(6), 1272-1281.e4.
Levine, J. (1983). Materialism and qualia: The explanatory gap. *Pacific Philosophical Quarterly*, 64(4), 354-361.
Nagel, T. (1974). What is it like to be a bat? Philosophical Review, 83(4), 435-450.
Oizumi, M., Albantakis, L., & Tononi, G. (2014). From the phenomenology to the mechanisms of consciousness: Integrated information theory 3.0. PLOS Computational Biology, 10(5), e1003588.
Parr, T., *et al.* (2022). Active Inference: The Free Energy Principle in Mind, Brain, and Behavior. MIT Press.
Pfeifer, R., & Bongard, J. (2006). *How the Body Shapes the Way We Think: A New View of Intelligence*. MIT Press.
Place, U. T. (1956). Is consciousness a brain process? *British Journal of Psychology*, 47(1), 44-50.
Rosenthal, D. M. (2005). *Consciousness and Mind*. Oxford University Press.
Ryle, G. (1949). *The Concept of Mind*. Hutchinson.
Seth, A. K., Suzuki, K., & Critchley, H. D. (2011). An interoceptive predictive coding model of conscious presence. *Frontiers in Psychology*, 2, 395.
Seth, A. K. (2013). Interoceptive inference, emotion, and the embodied self. *Trends in Cognitive Sciences*, 17(11), 565-573. https://doi.org/10.1016/j.tics.2013.09.007
Shannon, C. E. A mathematical theory of communication. *Bell System Technical Journal*, 27, 379-423, 623-656, (1948).
Simen, P. (2012). Evidence accumulator or decision threshold - which cortical mechanism are we observing? *Frontiers in Psychology*, 3, 183.
Smart, J. J. C. (1959). Sensations and brain processes. *Philosophical Review*, 68(2), 141-156.
Solms, M., & Panksepp, J. (2012). The Id knows more than the Ego admits: Neuropsychoanalytic and primal consciousness perspectives on the interface between affective and cognitive neuroscience. *Brain Sciences*, 2(2), 147-175.
Solms, M. (2021). The Hidden Spring: A Journey to the Source of Consciousness. Profile Books.
Starzyk, J. A. (2008). *Motivation in embodied intelligence*. In Frontiers in Robotics, Automation and Control (pp. 83-110). I-Tech Education and Publishing.
Starzyk, J. A., Raif, P., & Tan, A.-H. (2010). *Mental development and representation building through motivated learning*. WCCI 2010 - Special Session on Mental Architecture and Representation, Barcelona, Spain.
Starzyk, J. A. (2012). Motivated learning for computational intelligence. In *Machine Learning: Concepts, Methodologies, Tools and Applications* (pp. 120-146). IGI Global Scientific Publishing.
Starzyk, J. A., *et al.* (2012). Motivated learning for the development of autonomous systems. *Cognitive Systems Research*, 14(1), 10-25.
Starzyk, J. A., *et al.* (2013). *About motivation and cognition*. Manuscript.
Stevens, S. S. (1957). On the psychophysical law. *Psychological Review*, 64(3), 153-181.
Sutton, R. S., & Barto, A. G. (1998). *Reinforcement Learning: An Introduction*. MIT Press.
Tononi, G. (2004). An information integration theory of consciousness. *BMC Neuroscience*, 5, 42.
Wiener, N. (1948). *Cybernetics: Or Control and Communication in the Animal and the Machine*. MIT Press.
Wong, K.-F., & Wang, X.-J. (2007). Neural circuit dynamics underlying accumulation of time-varying evidence during perceptual decision making. *Frontiers in Computational Neuroscience*, 1, 6.